\documentclass[mathleft
]{an}
\usepackage{graphicx}
\usepackage{times}
\usepackage[authoryear]{natbib}
\begin{document}

% The following seven commands are intended for editorial usage and should be ignored by
% the author(s).
\Pagespan{789}{}% Document's page range. 
% If second parameter is left empty, the last page is computed automatically.
\Yearpublication{2009}%
\Yearsubmission{2009}%
\Month{}%   
\Volume{999}%  
\Issue{88}% 
% \DOI{This.is/not.aDOI}% 

\title{A study of major mergers using a multi-phase ISM code}

\author{J. Weniger\inst{1}\fnmsep\thanks{\email{julia.weniger@univie.ac.at}\newline}
%Example 
%for footnote, note the usage of the \texttt{fnmsep}
%command as separator between institute number and footnote mark} 
\and  Ch. Theis\inst{1} \and  S. Harfst\inst{2}
}
\titlerunning{A study of major mergers using a multi-phase ISM code}
\authorrunning{J. Weniger, Ch. Theis, \& S. Harfst}
\institute{
Institut f\"ur Astronomie, Universit\"at Wien, T\"urkenschanzstra\ss{}e 17, 1180 Wien, Austria
\and 
Leiden Observatory, Leiden University, PO Box 9513, 2300 RA Leiden, the 
Netherlands}

\received{}
\accepted{}
\publonline{}


\newcommand{\aap}{A{\textnormal\&}A}

\abstract{%    
Galaxy interactions are a common phenomenon in clusters of galaxies.
Especially major mergers are of particular importance, because they
can change the morphological type of galaxies. They have an impact
on the mass function of galaxies and they trigger star formation - the
main driver of the Galactic Matter Cycle. Therefore, we conducted a study of major
mergers by means of a multi-phase ISM code. This code is based on a TREE-SPH-code
combined with a sticky particle method allowing for star formation 
controlled by the properties of a multi-phase ISM. This is in contrast to the usually
implemented Schmidt law depending mainly on the gas density. Previously, this code was used on isolated
galaxies. Since our star formation recipe is not restricted to a special type of galaxy, it is interesting to apply it to interacting
galaxies, too. Our study on major mergers includes a research
of global properties of the interacting system, namely the star formation
rate and the star formation efficiency, the evaporation and condensation rates,
as well as the mass exchange of distinct components, namely stars, diffuse
ISM, and clouds. Investigating these properties provides insight to
interrelations between various physical processes. The results indicate that the star formation efficiency 
as well as the evaporation and condensation rates are influenced by the interaction.
}

\maketitle

\section{Introduction}

For some decades it has been known, that galaxy interactions play an important role in the evolution of galaxies changing their morphological type or triggering star formation.
Since it is possible to take remarkable pictures of peculiar galaxies 
(see e. g. \citealt{1966apg..book.....A}), the question was raised if 
their features are a result of galaxy interactions. Peculiar features
are e.g. bridges connecting galaxies or tails protruding from them.
The above mentioned assumption seemed likely, however, it was not clear until numerical
simulations corroborated it. The first simulations only considered
gravitational forces (e.g. \citealt{1941ApJ....94..385H}; Pleiderer \& Sieden\-topf 1961; \citealt{1972ApJ...178..623T}; \citealt{1988ApJ...331..699B}),
using different simplifications on the force calculation owing
to the limited computing power at that time. Nevertheless, tho\-se simulations already demonstrated, that interactions between galaxies change their appearance.

Simulations became more sophisticated with increasing computer power.
Hence, it was feasible to address topics concerning the interstellar
medium (ISM). Though the ISM only contributes approximately 10\% of the
mass of present-day galaxies, it is still an important ingredient:
firstly, all stars are formed within the cold phase of the ISM, the
mo\-lecular clouds. Hence, it is necessary to consider this component
in order to study the star formation rate during ga\-laxy interactions.
Secondly, the dynamics are strongly influenced by hydrodynamics and especially energy dissipation.
Which properties of the ISM are considered depends on the code used (see e.g. the review by \citealt{1992ARA&A..30..705B}).
Our code incorporates a 3-phase ISM similar to the proposition by \citet{1977ApJ...218..148M}.
The code was developed in order to
study Milky Way-like galaxies accounting for several components, namely
stars, dark matter, and the three phases of the ISM (\citealt{2006A&A...449..509H}). The warm and hot
phases are treated by smoothed particle hydrodynamics implemented similar to \citet{1989ApJS...70..419H}.
The molecular clouds are modelled by the sticky particle method (for details refer to \citealt{1993A&A...280...85T}).
A similar code was used by \citet{2002A&A...388..826S}, but they used a different sticky particle method and a different
star formation prescription.

Major mergers, that are interacting galaxies of about equal size, are
studied by numerous authors. Often the research's aim is to study
the properties of the remnant ga\-laxy (e.g. \citealt{1988ApJ...331..699B,2005ApJ...622L...9S}),
or to look for substructures, such as tidal dwarf galaxies (\citealt{2004A&A...427..803D}; \citealt{2006A&A...456..481B}; \citealt{2007MNRAS.375..805W}). Other
goals are the reproduction of observed major mergers (e.g. \citealt{2008AN....329.1042K}),
or to study the overall properties of interacting galaxies, such as the star formation rate (e.g. Mihos \& Hernquist 1996; \citealt{2008A&A...492...31D}).
While studying the star formation most often the Schmidt-Kennicutt
relation (\citealt{1959ApJ...129..243S}; \citealt{1998ApJ...498..541K}) or some modification
of it are used. The Schmidt-Kennicutt relation says that the star formation rate is proportional to the gas density, with the proportionality factor being a constant star formation efficiency. This work is distinct in that respect, since it does
not use the Schmidt-Kennicutt relation, but uses a star formation
prescription following \citet{1997ApJ...480..235E}. They found that
the star formation efficiency is not fixed, but dependent on local
properties of the ISM, i.e. the mass of the star forming cloud and the pressure in the ambient ISM.

The main aim of this work is to study the evolution of the star formation during a major merger. In particular we are interested in understanding how a locally defined star formation efficiency affects the star formation rate. Our study also yields the mass exchange rates due to condensation and evaporation.

\section{Interactions between different components}

\begin{figure}
\includegraphics[scale=0.45]{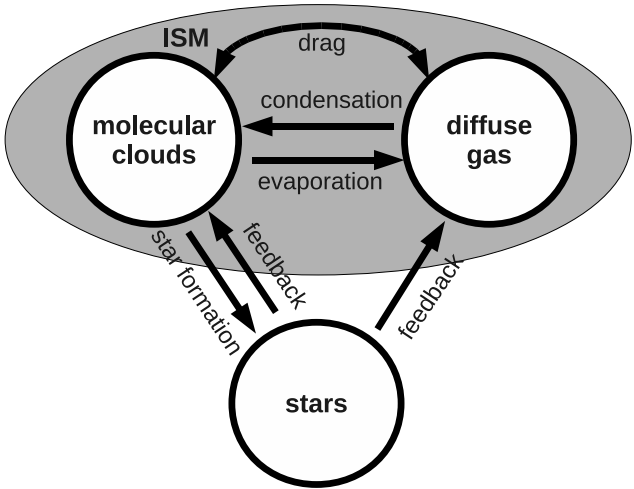}
\caption{\label{fig:physics}\small A scheme of all physical processes connecting 
the different components of a galaxy. In addition, the gas looses energy by radiative cooling and the molecular clouds by inelastic collisions.}
\end{figure}

The interaction participants are described by particles mimicking distinct components 
of a galaxy, namely stars, diffuse gas, molecular clouds, and dark matter particles.
Those components are connected by various physical processes (see Fig. \ref{fig:physics}).
The processes condensation, evaporation, and a drag force due to ram pressure allow
for matter and momentum exchange between the diffuse ISM and the clumpy
molecular clouds. Star formation and related feedback
of supernovae type II and planetary nebulae close the circuit of matter. Energy dissipation is caused by radiative cooling
in case of the diffuse ISM and by inelastic collisions in case
of molecular clouds.
Since studying the evolution of star formation is the main aim of this work, 
the next section will describe its implementation in more detail. For a 
thorough description of all other processes we refer to \citet{2006A&A...449..509H}.

\subsection{Star formation}

\begin{figure}
\includegraphics[scale=0.35]{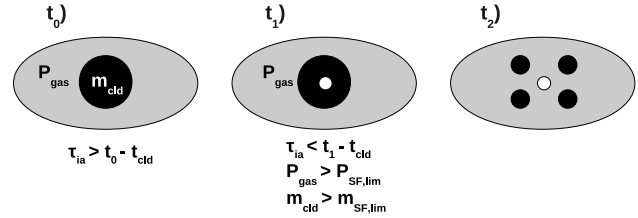}
\caption{\label{fig:starformation} \small The star formation scheme according to \citet{2006A&A...449..509H}. First the molecular cloud is inactive for a period $\tau_{\mathrm{ia}}$ ($t_{0}$). After that time span, a "star cluster" is formed within the molecular cloud provided that all other star formation criteria are fulfilled ($t_{1}$). The mass of the embedded star cluster is dependent on the local star formation efficiency, $\epsilon$. Finally, the cloud is fragmented due to energy input from supernovae type II and the ambient gas is heated ($t_{2}$).} 
\end{figure}

The site of star formation is known to be giant molecular clouds (e.g. \citealt{2003ARA&A..41...57L}). Since they are 
represented in our code by sticky particles, those particles have to 
be converted to star particles. The implementation of it works as follows (Fig. \ref{fig:starformation}):

\begin{itemize}
\item We assume that it takes $\tau_{\mathrm{ia}}=200\,\mathrm{Myr}$ till gas in a molecular cloud is able to form stars. During this time the molecular cloud is inactive. In other words, star formation is suppressed as long as the age of the molecular cloud, $t_{\mathrm{now}}-t_{\mathrm{cld}}$ is smaller than $\tau_{\mathrm{ia}}$. $\tau_{\mathrm{ia}}$ is the only free parameter in our star formation prescription and can be interpreted as a global star formation time scale. It is gauged by the star formation rate of the Milky Way. Since the mean star formation efficiency, $\epsilon$, in our simulations is approximately 5\% (see Fig. \ref{fig:evolution}) and the total mass in molecular clouds is approximately $4\cdot10^{11}\mathrm{M}_{\odot}$ (see Table \ref{tab:Mass-and-number}), it gives a mean star formation rate of approximately $\mathrm{SFR}=\tau_{\mathrm{ia}}^{-1}\cdot\epsilon\cdot\mathrm{M_{cld}}\approx1\,\mathrm{M}_{\odot}\,\mathrm{yr^{-1}}$.
\item Afterwards stars can be formed. They stay embedded in their parent molecular cloud.
Their mass is determined by the star formation efficiency, $\epsilon$. A mass criterion and a pressure criterion prevent the 
formation of too small stellar particles: the mass of the molecular clouds, $m_{\mathrm{cld}}$, must exceed $2.5\cdot10^{4}\mathrm{M}_{\odot}$ and the pressure at the position of the molecular cloud, $P_{\mathrm{gas}}$, must be higher than a tenth of the ISM pressure at the position of the Sun in the Milky Way, i.e. $P_{\odot}=3\cdot10^{4}\mathrm{K\, cm^{-3}}$ (Elmegreen \& Efremov 1997). Note, that the star formation efficiency is not constant, but it depends on local properties, namely the gas pressure and the mass of the parent molecular cloud.
\item After the most massive stars end their lives as supernovae, the molecular cloud fragments into four pieces of equal mass due to feedback by supernovae.
\end{itemize}

\section{Numerical setup}

Following \citet{2006A&A...449..509H} the initial galaxies were constructed
by firstly generating a disk/bulge/halo - system using the method
of \citet{1995MNRAS.277.1341K}. In the next step, one fifth of the
stellar disk particles were transformed into molecular clouds and finally
the diffuse ISM was added as an initially slowly rotating homogeneous
sphere, which will collapse in the first 200 Myr forming a warm disk. The collisionless model of \citet{1995MNRAS.277.1341K} realises an equilibrium configuration. 
However, the equilibrium is affected by adding the ISM and the
above mentioned processes. Therefore, we follow the system's evolution, until a quasi-equilibrium is established. The numerical integration
is done by means of a TREE-SPH code combined with the sticky particle method (\citealt{1993A&A...280...85T}). 
Gravitational forces are determined by the DEHNEN-Tree (\citealt{2002JCoPh.179...27D}). 
See \citet{2006A&A...449..509H} for a more thorough description.

The relaxed galaxy has a mass of $2.46\cdot10^{11}\,\mathrm{M}_{\odot}$,
thereof 79\% in dark matter particles. The remaining baryonic matter
consists of approximately 12\% ISM particles - hence, molecular clouds
and diffuse ISM - the rest being disk and bulge stars. The galaxy
is realised by a total number of 230462 particles. Note, that not only the mass,
but also the number of clouds, SPH-particles, and disk stars as given
in Table \ref{tab:Mass-and-number} will change throughout the simulations:
this is caused by inelastic collisions, star formation, and feedback. 
The numbers of dark matter particles ($10^{5}$) and bulge stars ($10^{4}$), on the
other hand, stay constant throughout the simulations. Both galaxies are of equal mass and their rotation is prograde.
The galactic planes coincide with the orbital plane. At $t=0$ the
galaxies are placed at a separation of 100 kpc. Their initial speed
is chosen to match a parabolic orbit with a minimum
separation of 20 kpc. The run was stopped at $t=3\,\mathrm{Gyr}$.

\begin{table}
\caption{\label{tab:Mass-and-number}Mass of distinct components of one galaxy
and corresponding number of particles}
\begin{tabular}{rrr}
\hline 
Component & Mass {[}$10^{11}\mathrm{M}_{\odot}${]} & NR. of particles\tabularnewline
\hline
\hline 
Bulge-stars & 0.17 & 10000\tabularnewline
Disk-stars & 0.29 & 75098\tabularnewline
Clouds & 0.04 & 36347\tabularnewline
Gas-particles & 0.02 & 9017\tabularnewline
DM-particles & 1.94 & 100000\tabularnewline
\hline 
Total & 2.46 & 230462\tabularnewline
\hline
\end{tabular}
\end{table}

\section{Results}

\begin{figure*}
\includegraphics[scale=0.25]{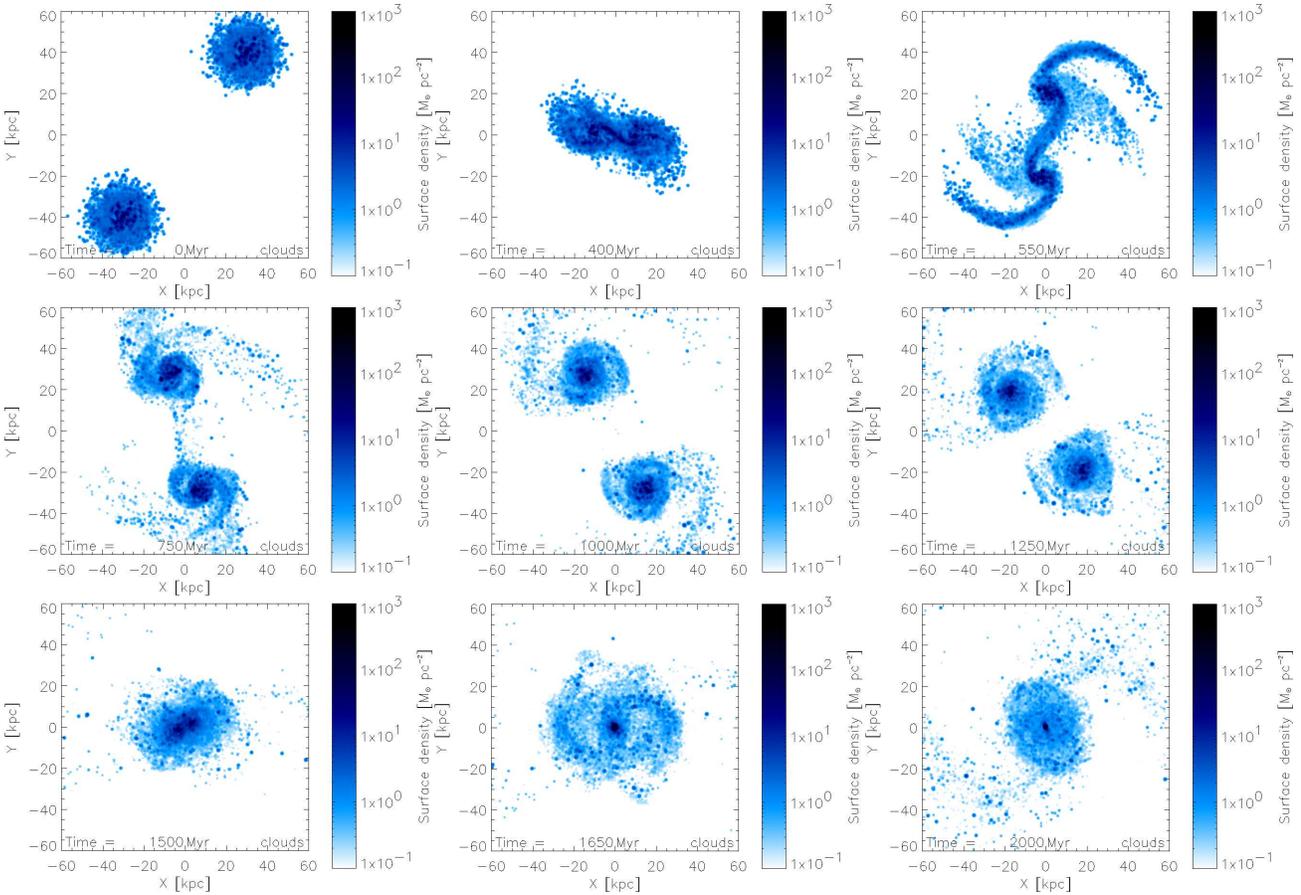}\caption{\label{fig:interaction} \small Snapshots of the interaction. Surface density maps of the clouds are shown (from left to right and top to bottom: t equals 0, 400, 550, 750, 1000, 1250, 1500, 1650, 2000 Myr). The first passage takes place at t = 400 Myr, the second at t = 1520 Myr, and the merging starts at t = 1650 Myr.}
\end{figure*}

\begin{figure*}
\includegraphics[scale=0.42]{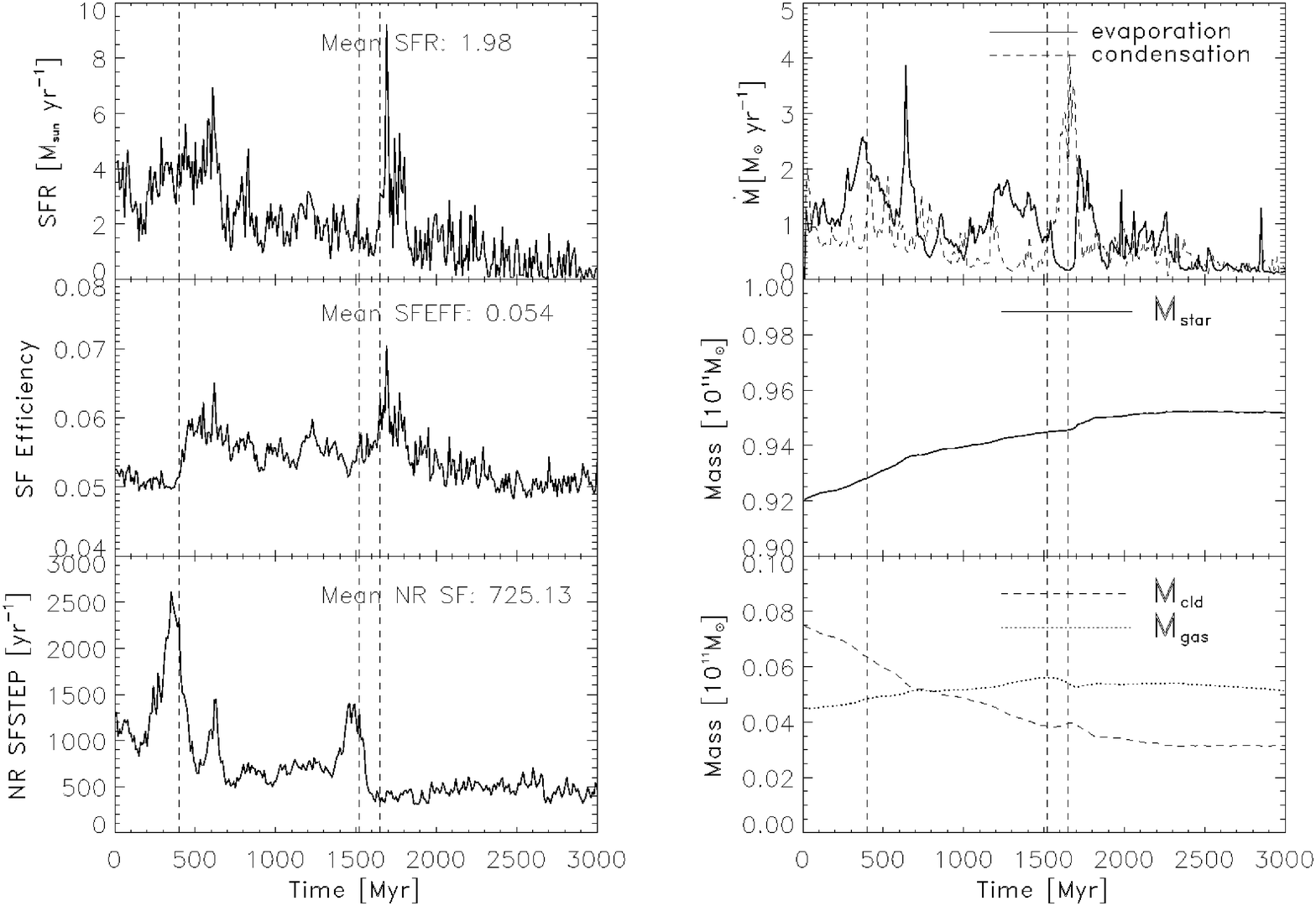}\caption{\label{fig:evolution} \small Evolution of star formation (left) and mass (right): the star formation rate (left, top), the evolution of the mean star formation efficiency (left, middle), and the number of star formation processes within a sampling period of 10 Myr (left, bottom). On the right side, the evolution of the evaporation and condensation rates (top), the evolution of mass of stars (middle), and 
the evolution of mass of the ISM (bottom) are plotted. The three vertical dashed lines indicate the first passage, 
the second passage, and the beginning of the merging.}
\end{figure*}

\begin{figure}
\includegraphics[scale=0.33]{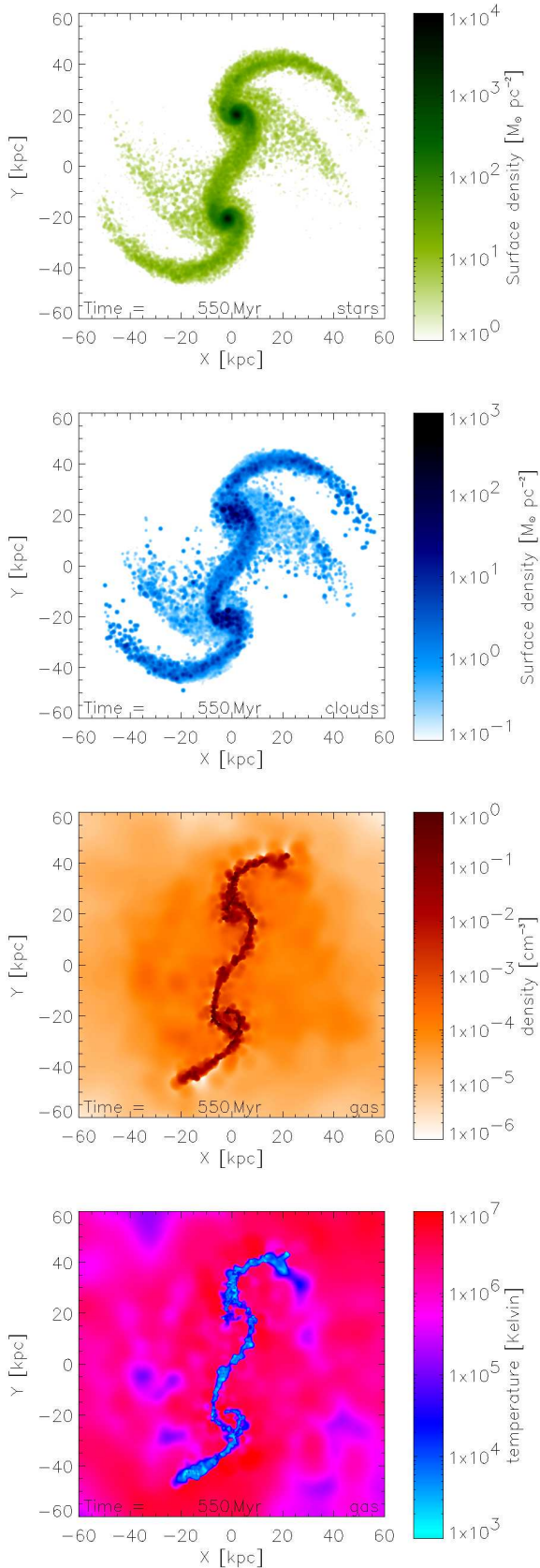}
\caption{\label{fig:snapshot} \small This figure shows from top to bottom a surface density map of stars, 
a surface density map of clouds, a density map of the diffuse ISM, and a temperature map of the diffuse 
ISM at time $t=550\,\mathrm{Myr}$ (i.e. 150 Myr after the first passage).}
\end{figure}

The code configuration allows to follow the evolution of different components and their interrelations. To illustrate the time evolution, a series of snapshots are plotted in Fig. \ref{fig:interaction} exhibiting the surface density of the cold phase. The distribution of the cold phase resembles the distribution of the stars very well. This implies that dissipation does not affect the cold phase severely. In contrast, the diffuse ISM shows a slightly different distribution (see also Fig. \ref{fig:snapshot}). The fist snapshot of Fig. \ref{fig:interaction} (top, left) shows the initial configuration. The first passage at $t=400\,\mathrm{Myr}$ is pictured in the second panel (top, middle). Afterwards, a series of snapshots features on the one hand the expansion of the tails and on the other hand the development of a bridge, which vanishes gradually. The bottom row of Fig. \ref{fig:interaction} shows a snapshot shortly before the second passage which takes place at $t=1520\,\mathrm{Myr}$, the beginning of the merging at $t=1650\,\mathrm{Myr}$, and a quiescent phase after the merging ($t=2000\,\mathrm{Myr}$). The 
beginning of the merging is defined by the last local minimum in the 
potential energy of the system, after which the centre of mass of the 
bulge components are no longer separating.
In addition, Fig. \ref{fig:evolution} shows the star formation rate, the star
formation efficiency, the number of star formation processes per 10 Myr, the evaporation and condensation rate
and the evolution of the masses of the different components. The vertical dashed lines 
in Fig. \ref{fig:evolution} indicate the first encounter, the second encounter, and the 
beginning of the merging.

\begin{figure}
\includegraphics[scale=0.35]{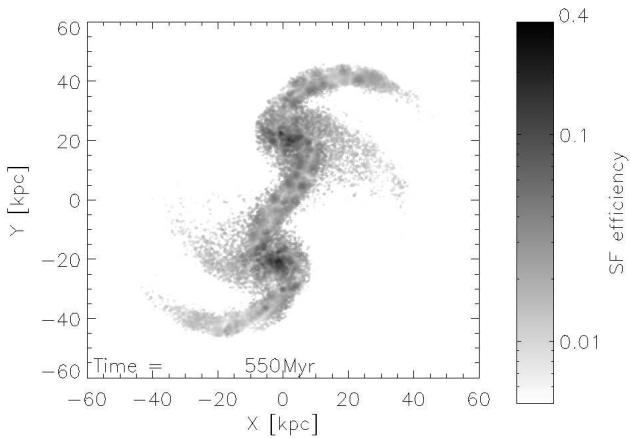}

\caption{\label{fig:sfeff} \small The star formation efficiency at $t=550\,\mathrm{Myr}$.}

\end{figure}

\begin{figure}
\includegraphics[scale=0.33]{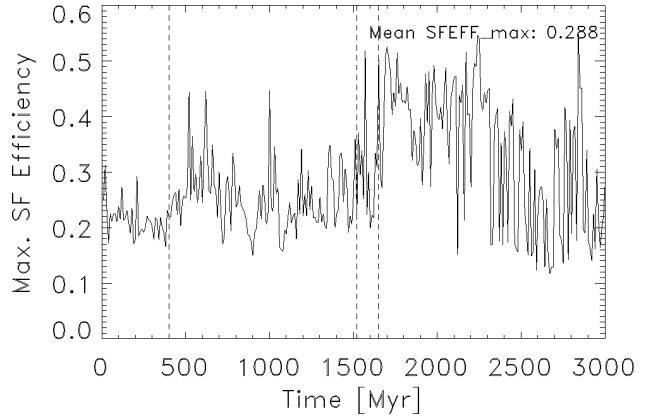}

\caption{\label{fig:sfeffmax} \small Evolution of the maximum star formation efficiency. The three vertical dashed lines indicate the first passage, 
the second passage, and the beginning of the merging.}

\end{figure}

The star formation rate shows fluctuations due to the limited
number of star formation processes during measurement intervals of 10 Myr. Still it is obvious that there is a maximum in the star formation rate at 1690 Myr, shortly
after the merging starts. Another local maximum is given at 610 Myr,
that is 210 Myr after the first passage. However, Fig. \ref{fig:evolution} 
suggests an enhanced star formation already prior to the first passage. 
The period of enhanced
star formation located around the first passage is more than 400 Myr long. 
In this setup, the
reason for enhanced star formation is not an enhanced
number of star formation processes, but an enhanced star formation
efficiency: the number of star formation processes does not correlate
well with the star formation rate, but the star formation efficiency does,
as one can easily infer from the left side of Fig. \ref{fig:evolution}. Only the enhanced star formation prior to the first encounter cannot be explained by an enhanced star formation efficiency. Here, it seems that the number of 
star formation processes is crucial. 
Although the mean star formation efficiency is only 5\%, the maximum star formation efficiency found within a time step can even exceed 50\%. In general, Fig. \ref{fig:sfeffmax} suggests, that the merging facilitates higher star formation efficiencies. In fact, the maximum star formation efficiency is in the mean about 9\% higher after $t=1650\,\mathrm{Myr}$. Fig. \ref{fig:sfeff} features a map of the star formation efficiency at $t=550\,\mathrm{Myr}$, 
hence 150 Myr after the first passage. The star formation efficiency is highest 
in the centre of the galaxies, but also sites of enhanced star formation efficiency  are recognised 
in the bridge and the tails.
Though the star formation rate is affected by the interaction, it is still not extraordinary high. 
Previous research (e.g. \citealt{2004ApJ...607L..87C}) implies that the magnitude of the star formation rate depends on the initial conditions, especially on 
the orbit. Furthermore the fraction of mass in molecular clouds might be too low to 
obtain a starburst of several tens up to several hundreds $\mathrm{M}_{\odot}\,\mathrm{yr^{-1}}$.

Other important quantities are 
the evaporation and condensation rates. The local maxima in the
evaporation rate (at $t=630\,\mathrm{Myr}$ and at $t=1710\,\mathrm{Myr}$) take place approximately 20 Myr after the local maxima
in the star formation rate. This time interval corresponds to the time delay between star formation and feedback due to energy
input of supernovae. Note, however, that the time resolution is only
10 Myr. The condensation rate has its maximum at exactly the same time the merging starts.
During this episode condensation outweighs evaporation. In general, however, evaporation exceeds condensation. The mean value
for the former is $1\,\mathrm{M}_{\odot}\,\mathrm{yr^{-1}}$ and for the latter $0.8\,\mathrm{M}_{\odot}\,\mathrm{yr^{-1}}$. A high condensation rate 
coinciding with the beginning of the mer\-ging, 
has also been found in other major mergers performed by us so far. The understanding 
of this remarkable feature and its implications will be the topic of 
further study.

Altogether the mass evolution of distinct components shows the following features 
(see middle and bottom diagram of Fig. \ref{fig:evolution})

\begin{itemize}
\item In general, the cloud mass is depleted by star formation while the stellar mass increases continuously. Furthermore, the cloud mass decreases, because evaporation outweighs condensation.  As a result only 3\% of baryonic matter are molecular clouds in the end of the simulation. In comparison, the fraction of molecular clouds was initially 8\%.
\item Due to star formation and related feedback during the first passage the stellar mass as
well as the mass of the diffuse ISM rises more steeply.
\item As a consequence of the merging accompanied by enhanced star formation, the stellar mass increases rapidly. In Fig. \ref{fig:evolution} one can also distinguish a short period of time, within which the diffuse ISM is diminished by condensation, whereas the cloud mass increases.
\end{itemize}

Fig. \ref{fig:snapshot} shows  a snapshot of the interaction at t = 550 Myr - 150 Myr after the first passage. Different components are plotted separately in order to emphasise the multi-phase properties of our simulation. In the uppermost panel a surface density map of stars is shown. It is easy to pinpoint the two galaxies, a bridge connecting them, and tails protruding from each of them. The cold molecular clouds (second panel from top) follow the distribution of the stars very well. However, the tail of the warm phase is considerably shorter as one can deduce from the third and forth panel which show a density map of the diffuse ISM and a temperature map, respectively. In the temperature map it is easy to identify two phases: a warm phase in the galactic disks, the bridge, and the tails, and a hot phase primarily located in the halo, but also in hot supernovae bubbles. The reason behind the shortened tails is the initial setup of the diffuse gas component. Due to the initial collapse, the warm gaseous disk is initially much smaller than the 
stellar disk and grows only slowly in size. At the beginning of the interaction the warm gaseous disk is still 
smaller than the stellar disk. Since the utmost part of the tail evolves from the utmost part of the disk, the tail of the warm ISM is not as extended as the stellar tail or the tail of the cold phase.

\section{Conclusions \& Outlook}

We have presented a study of an equal size major merger by means of 
a multi-phase ISM code described in \citet{2006A&A...449..509H}. We have focused 
on the evolution of the star formation rate as well as on the evaporation and 
condensation rates. The multi-phase nature of the code allows for an extensive analysis 
of mass transfer involving galaxy interactions. Furthermore, the star formation 
prescription according to \citet{1997ApJ...480..235E} allows to follow the star formation efficiency during the interaction locally. The star formation rate as well as the star formation efficiency increases due to the first encounter and the merging of the nuclei. In the mean the star formation efficiency is only 5\%, but at some locations, usually in the centre of the galaxies, but also in the bridge or the tails, the star formation efficiency can be much higher - up to 55\%.
Yet, a drawback of our models is an initial gaseous disk which is smaller than the stellar disk resulting in a lack of SPH particles at the tip of the tails. Therefore, an improved setup of the SPH component is needed in order to study substructures, such as tidal dwarf galaxies, within the tip of the tails.

\acknowledgement
JW is a member of the Initiativkolleg (IK) 'The Cosmic Matter Circuit' I033-N of the University of Vienna. 
The numerical simulations were performed on the Grape cluster of the Institute of Astronomy of the University of Vienna. CT is grateful for financial support by the project TH511/9-1 within the DFG Priority Programme 1177 'Galaxy Evolution'. SH is supported by the NWO Computational Science STARE project 643200503.

\end{document}